\documentclass[runningheads]{svmult}

\usepackage{makeidx}   
\usepackage{graphicx}  
\usepackage{subeqnar}  
\usepackage{multicol}  
\usepackage{physprbb}  
\makeindex             

\newcommand{\etal}{{et al.~}}

\newcommand{\kms}{\>{\rm km}\,{\rm s}^{-1}}

\newcommand{\kpc}{\>{\rm kpc}}
\newcommand{\Msun}{\>{\rm M_{\odot}}}

\begin{document}
\title*{Measuring the Virial Masses of Disk Galaxies}
\toctitle{Measuring the Virial Masses of Disk Galaxies}

%
%
\titlerunning{Measuring the Virial Masses of Disk Galaxies}
%
\author{Frank C. van den Bosch}
\authorrunning{Frank C. van den Bosch}
%
%
\institute{Max-Planck Institut f\"ur Astrophysik, Postfach 1317, 85741 Garching, Germany}

\maketitle              

\begin{abstract}
  I present  detailed models for  the formation of disk  galaxies, and
  investigate  which  observables  are  best  suited  as  virial  mass
  estimators.   Contrary to naive  expectations, the  luminosities and
  circular velocities  of disk galaxies are  extremely poor indicators
  of  total virial mass.   Instead, I  show that  the product  of disk
  scale length and rotation velocity squared yields a much more robust
  estimate.  Finally, I show  how this  estimator may  be used  to put
  limits  on  the efficiencies  of  cooling  and  feedback during  the
  process of galaxy formation.
\end{abstract}

\section{Introduction}

Currently, the  main uncertainties in our picture  of galaxy formation
are related to the intricate processes of cooling, star formation, and
feedback.   The  cooling  and  feedback  efficiencies  are  ultimately
responsible  for setting  the  galaxy mass  fractions  $f_{\rm gal}  =
M_{\rm  gal} / M_{\rm  vir}$.  Here  $M_{\rm gal}$  is the  total {\it
baryonic} mass of the galaxy (stars plus gas, excluding the hot gas in
the halo) and $M_{\rm vir}$ is the total virial mass.

Here I present new models for  the formation of disk galaxies, which I
use to  investigate how well observables {\it  extracted directly from
these models} can be used to recover $f_{\rm gal}(M_{\rm vir})$.  Even
though  the  assumptions  underlying  the model  are  not  necessarily
correct, and  the phenomenological descriptions of  star formation and
feedback are  certainly oversimplified, this  provides useful insights
regarding the  ability of actual observations to  constrain the poorly
understood astrophysical processes of galaxy formation.

\section{Short Description of Models}

The main assumptions that characterize the framework of the models are
the following: (i) dark matter  halos around disk galaxies grow by the
smooth accretion of  mass, (ii) in the absence  of cooling the baryons
have the  same distribution of mass  and angular momentum  as the dark
matter, and (iii) the baryons conserve their specific angular momentum
when they cool.   I follow Firmani \& Avila-Reese  (2000) and make the
additional assumptions that (iv) the  spin parameter of a given galaxy
is constant with time, (v) each mass shell that virializes is in solid
body rotation, and  (vi) the rotation axes of  all shells are aligned.
Although neither of these  assumptions is necessarily accurate, it was
shown  in  van den  Bosch  (2001) that  they  result  in halo  angular
momentum  profiles in  excellent  agreement with  the high  resolution
$N$-body simulations of Bullock et al. (2001).

The main outline of the models is  as follows.  I set up a radial grid
between $r=0$  and the present day  virial radius of  the model galaxy
and follow the formation and evolution  of the disk galaxy using a few
hundred time steps.  Six  mass components are considered: dark matter,
hot gas,  disk mass (both in stars  and in cold gas),  bulge mass, and
mass ejected by outflows from the disk.  The dark matter, hot gas, and
bulge mass are assumed to  be distributed in spherical shells, whereas
the disk stars and cold gas  are assumed to be in infinitesimally thin
annuli.  Each time  step the changes in these  various mass components
in each radial bin are computed using the following prescriptions.
\begin{itemize}

\item The rate at which the total virial mass grows with time is given
by the Universal mass accretion history (van den Bosch 2002a).

\item At each redshift, the dark matter is assumed to follow an NFW
density distribution (Navarro, Frenk \& White 1997).

\item The gas that enters the virial  radius of a halo is added to the
disk  a time $t_c  \equiv {\rm  max}[t_{\rm ff},t_{\rm  cool}]$ later,
where $t_{\rm ff}$ and $t_{\rm cool}$ correspond to the free-fall time
and cooling time.  The cooling time depends on  the metallicity of the
hot gas, which is taken to be a free parameter $Z_{\rm hot}$.

\item The radius at which the  gas settles is governed by its specific
angular  momentum  distribution, which  follows  from the  assumptions
(iv)--(vi) listed above.

\item When  the disk  becomes unstable, part  of the disk  material is
converted into a bulge component (cf. van den Bosch 1998).

\item In the disk, only the  cold gas with a surface density above the
critical  density  given by  Toomre's stability criterion  is
considered eligible  for star formation. This gas  is transformed into
stars with a rate given by a simple Schmidt law.

\item Part of  the cold gas is expelled from  the system by supernovae
feedback. This is  modeled as in van den  Bosch (2000), and regulated
by a free feedback efficiency parameter $\epsilon_{\rm fb}$.

\end{itemize}

{\noindent More details regarding these models can be found in van den
Bosch (2001), where it is  shown that these models yield disk galaxies
in good agreement with observations. For instance, for the majority of
the model galaxies the  disk reveals an exponential surface brightness
profile. Note  that contrary to previous disk  formation models (e.g.,
Mo, Mao  \& White 1998), this is  not an {\it a  priori} assumption of
the model.}
\begin{figure}[b]
\begin{center}
\includegraphics[width=.9\textwidth,clip=]{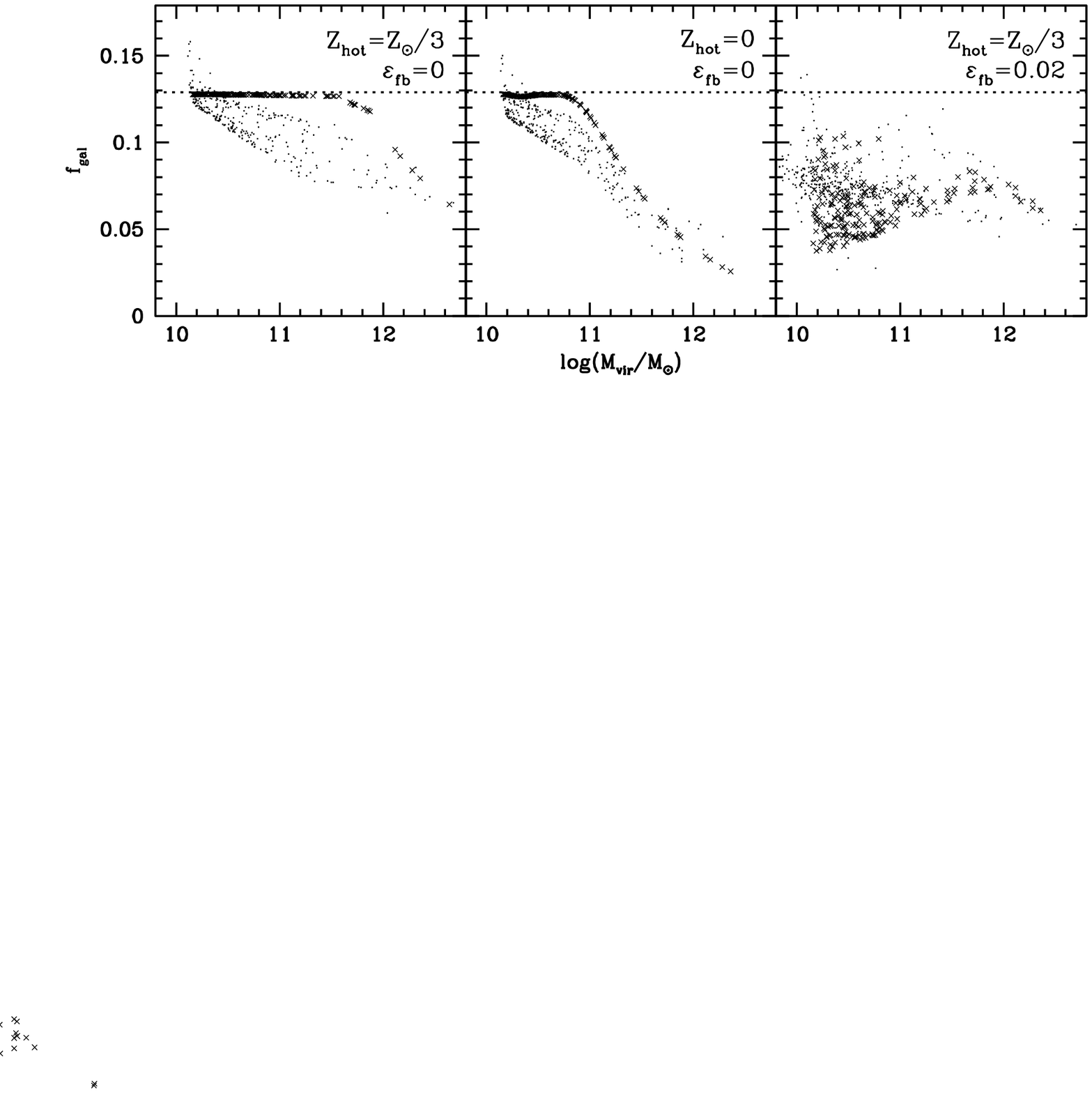}
\end{center}
\caption[]{A comparison of the true galaxy mass fraction $f_{\rm gal}$
as function  of the  true virial mass  (crosses) with the  same values
estimated  from  the observables  extracted  from  the models  (dots).
Results are plotted  for all three models discussed  in the text.  The
horizontal  dotted line  corresponds  to the  universal baryonic  mass
fraction $f_{\rm bar}$.  In the  model with feedback (right panel) the
dots occupy the  same parameter space as the  crosses, indicating that
the observables  allow one to  recover $f_{\rm gal}(M_{\rm  vir})$, at
least  in a  statistical sense.   In the  two models  without feedback
there  are significant  errors  in the  recovered $f_{\rm  gal}(M_{\rm
vir})$.   Yet, the  two  $f_{\rm gal}(M_{\rm  vir})$ are  sufficiently
different to  discriminate between the two models;  in particular, the
estimated galaxy  mass fractions nicely avoid the  upper right regions
of  parameter   space  which   contain  information  on   the  cooling
efficiencies.  Furthermore,   the  dots  occupy   different  areas  of
parameter space in  models with and without feedback,  such that there
is hope that  the ``observed'' $f_{\rm gal}(M_{\rm vir})$  may be used
to constrain the efficiency of feedback.}
\label{eps1}
\end{figure}

\section{Galaxy Mass Fractions}

In order to investigate how  $f_{\rm gal}(M_{\rm vir})$ relates to the
cooling  and feedback efficiencies  I discuss  three models  that only
differ  in the  metallicity of  the hot  gas, $Z_{\rm  hot}$,  and the
feedback efficiency, $\varepsilon_{\rm fb}$: the `Standard Model' with
$Z_{\rm hot}  = 0.3  Z_{\odot}$ (a  typical value for  the hot  gas in
clusters) and $\varepsilon_{\rm fb}=0$  (i.e., no feedback), the `Zero
Metallicity  Model' with  $Z_{\rm  hot} =  0.0$ and  $\varepsilon_{\rm
fb}=0$, and  the `Feedback Model'  with $Z_{\rm hot} =  0.3 Z_{\odot}$
and $\varepsilon_{\rm fb}=0.02$ (i.e., two percent of the SN energy is
converted  to kinetic energy).   All other  model parameters  are kept
fixed at  their fiducial  values (see van  den Bosch 2001).   For each
model a  sample of $400$  model galaxies is constructed.   Present day
virial masses  are drawn from  the Press-Schechter mass  function with
$10^{10} h^{-1} \Msun \leq  M_{\rm vir}(0) \leq 10^{13} h^{-1} \Msun$,
corresponding to $31 \kms \leq V_{\rm vir} \leq 312 \kms$, roughly the
range expected for galaxies.   Spin parameters, which parameterize the
specific  angular  momentum,  are  drawn  from  a  typical  log-normal
distribution.
\begin{figure}[b]
\begin{center}
\includegraphics[width=.98\textwidth,clip=]{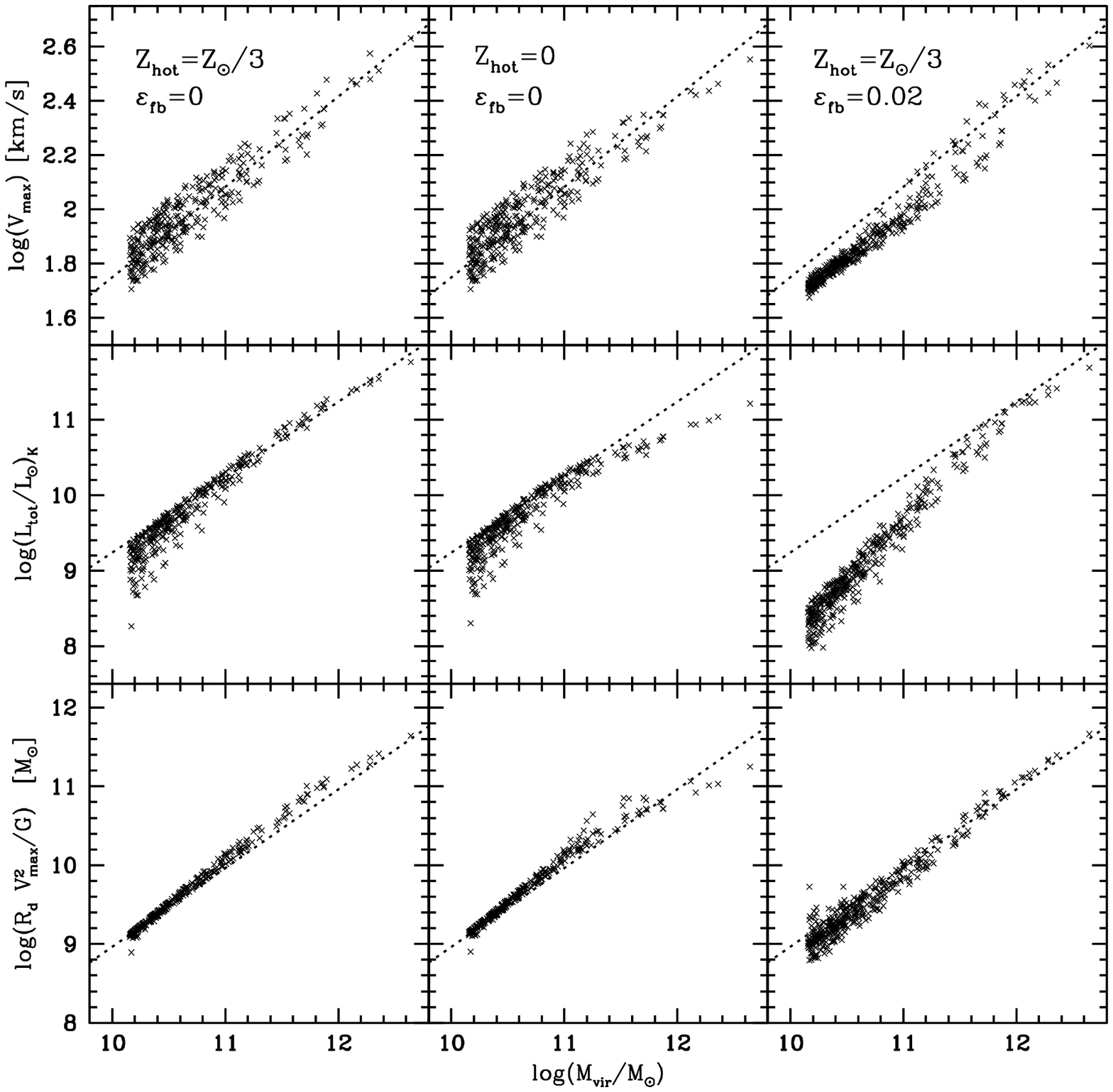}
\end{center}
\caption[]{The relation between various virial mass estimators and the
actual virial  mass for all three  models.  In the  upper panels ${\rm
log}(V_{\rm max})$  is plotted  versus ${\rm log}(M_{\rm  vir})$.  The
dotted line  corresponds to  $M_{\rm vir} \propto  V^{1/3}_{\rm max}$,
which is  a reasonable description of the  average relation.  However,
the scatter is large, and  the zero-point depends on the actual model,
which  makes $V_{\rm max}$  unsuitable as  virial mass  indicator. The
same goes for the $K$-band  luminosity, which is plotted in the middle
row of panels. Here the dotted line corresponds to $L_K \propto M_{\rm
vir}$,  which only yields  a reasonable  description for  the brighter
galaxies  in models  with  $Z_{\rm hot}  =  Z_{\odot}/3$. For  fainter
galaxies, however, the scatter is large.  Furthermore the slope of the
$L_K(V_{\rm   vir})$  relation  depends   strongly  on   the  feedback
efficiency.  Thus total  luminosity is also a poor  indicator of total
virial mass.   The lower  panels plot the  virial mass  estimator $R_d
V^2_{\rm  max} / G$  as function  of $M_{\rm  vir}$.  Here  the dotted
lines  correspond to equation~(\ref{mvir}),  which gives  a reasonable
fit to the model galaxies,  independent of the cooling and/or feedback
efficiencies. In addition, the scatter is relatively small, making the
product of disk  scale length and maximum rotation  velocity squared a
fairly accurate estimator of total virial mass.}
\label{eps2}
\end{figure}

Figure~1 plots  the present day  galaxy mass fractions $f_{\rm  gal} =
M_{\rm gal}/M_{\rm  vir}$ as function  of $M_{\rm vir}$  (crosses). In
the models  without feedback $f_{\rm  gal}$ is virtually  identical to
the universal baryon fraction $f_{\rm  bar}$ for low mass systems. For
more  massive systems,  cooling becomes  inefficient,  causing $f_{\rm
gal}$ to strongly decrease with  increasing virial mass.  This is more
pronounced in  the model  with $Z_{\rm hot}=0$,  for which  cooling is
least efficient. In the Feedback Model (right panel), $f_{\rm gal} \ll
f_{\rm bar}$  for the  low mass  systems, but with  a large  amount of
scatter.  This is a reflection of the scatter in halo spin parameters:
systems with  less angular momentum produce disks  with higher surface
densities, therefore have higher  star formation rates, which induce a
more efficient feedback. At the  high mass end $f_{\rm gal}$ is fairly
similar to the Standard Model without feedback.  This owes to the fact
that the mass ejection efficiency  scales inversely with the square of
the escape  velocity, making feedback  less efficient in  more massive
systems.

Clearly  $f_{\rm  gal}(M_{\rm  vir})$  depends strongly  on  both  the
cooling and  feedback efficiencies. Therefore,  if one could  obtain a
measure of  $f_{\rm gal}(M_{\rm vir})$ observationally  it would allow
us to constrain the poorly understood physics of cooling and feedback.
This requires  one to be able  to infer both the  baryonic galaxy mass
$M_{\rm  gal}$ as well  as the  total virial  mass $M_{\rm  vir}$ from
observations  of the  luminous  (and gaseous)  components.  Using  the
models  outlined above I  now investigate  which observables  are best
suited as the appropriate mass indicators.

For the  galaxy mass  one can  write $M_{\rm gal}  = \Upsilon_B  L_B +
M_{\rm  gas}$.    Here  $L_B$   is  the  total   $B$-band  luminosity,
$\Upsilon_B$  is the  corresponding stellar  mass-to-light  ratio, and
$M_{\rm  gas}$  is the  galaxy's  (cold)  gas  mass. $L_B$  is  easily
obtained  observationally,   and  I  assume  that   $M_{\rm  gas}$  is
observationally  accessible  through  HI  measurements.   The  stellar
mass-to-light  ratio,  however,  is  not  directly  accessible  to  an
observer,  but has  been shown  to correlate  strongly with  color.  I
therefore  extract  the $B-K$  color  from  the  models from  which  I
estimate $\Upsilon_B$ using the relation of Bell \& de Jong (2001).

The galaxy virial mass is more difficult to obtain. In Figure~2 I plot
three `observables' as function of  $M_{\rm vir}$ for galaxies in each
of the  three models.   Note that both  $V_{\rm max}$, defined  as the
maximum rotation velocity inside the  radial extent probed by the cold
gas, and  $L_K$ are poor indicators  of virial mass. First  of all the
slope  and   zero-points  of   the  $V_{\rm  max}(M_{\rm   vir})$  and
$L_K(M_{\rm  vir})$ relations depend  on the  input parameters  of the
model. This means that an  observer trying to infer $M_{\rm vir}$ from
either  $V_{\rm max}$  or $L_K$  needs to  make assumptions  about the
efficiencies  of cooling and  feedback. However,  it is  exactly these
efficiencies that we seek to constrain.  Secondly, the scatter of both
relations  can be  so  large that  even  if the  normalization of  the
relation were known, one could still not infer $M_{\rm vir}$ to better
than an  order of  magnitude.  In particular,  the scatter  in $V_{\rm
max}(M_{\rm  vir})$ can  be  very  large. This  owes  entirely to  the
scatter in  halo spin parameter,  which sets the concentration  of the
baryonic mass component  after cooling, therewith strongly influencing
$V_{\rm max}$.

As the lower panels in Figure~2 show, a much more reliable virial mass
indicator is  $R_d V_{\rm max}^2  / G$. Here  $R_d$ is the  disk scale
length in  the $I$-band, obtained  from fitting an exponential  to the
$I$-band surface brightness distribution of the disk. Upon fitting all
galaxies of all three models simultaneously I obtain
\begin{equation}
\label{mvir}
M_{\rm vir} = 2.54 \times 10^{10} \Msun \left({R_d \over \kpc}\right)
\left({V_{\rm max} \over 100 \kms}\right)^2.
\end{equation}
(rms scatter  between 20  and 50 percent,  depending on the  amount of
feedback).     The   fraction    of   model    galaxies    for   which
equation~(\ref{mvir}) yields  an estimate of  the true virial  mass to
better than a factor two is  larger than 97 percent!  It is remarkable
that the zero-point for models with feedback is the same as for models
without feedback.  When matter is ejected it reduces $V_{\rm max}$ but
at the same time increases the disk scale length such that $R_d V_{\rm
max}^2$ stays  roughly constant.   This is due  to the  star formation
threshold criterion included in the  models. If real galaxies follow a
similar threshold  criterion, equation~(1) provides  a fairly accurate
estimate of the total virial mass of disk galaxies.

We  now have all  the tools  in place  to see  whether we  can recover
$f_{\rm  gal}(M_{\rm vir})$  from  the ``observables''.   The dots  in
Figure~1  correspond to  the estimates  of $f_{\rm  gal}(M_{\rm vir})$
obtained  using the  method outlined  above.  In  both  models without
feedback  there  are  significant  errors  in  the  recovered  $f_{\rm
gal}(M_{\rm vir})$,  which is dominated  by errors in the  estimate of
$M_{\rm vir}$.   Yet, the recovered $f_{\rm gal}(M_{\rm  vir})$ of the
two models are sufficiently  different to distinguish between them. In
particular, in  both models  the dots avoid  the regions in  the upper
right  corner  which  contains  information about  the  efficiency  of
cooling.  In  the feedback model  the recovered values  occupy roughly
the  same  area of  the  $f_{\rm  gal} -  M_{\rm  vir}$  plane as  the
intrinsic   values.   Although   the  one-to-one   correspondence  for
individual  model galaxies may  be poor,  statistically the  method to
recover  $f_{\rm  gal}(M_{\rm vir})$  explored  here works  reasonably
well.

\section{Conclusions}

We  have shown  that  the product  of  disk scale  length and  maximum
rotation velocity squared  can be used as a  fairly accurate estimator
of total virial mass. This in  turn can be used to obtain estimates of
the galaxy mass  fractions as function of virial  mass, which contains
important information about the  efficiencies of cooling and feedback.
Another approach, that has been taken in the past, is to use published
luminosity  functions and  luminosity-velocity relations  to construct
halo velocity  functions (i.e., Newman  \& Davis 2000;  Gonzales \etal
2000; Kochanek  2001).  The main goal  of these studies  is similar to
the work presented here, namely to circumvent the problems with poorly
understood   astrophysical  processes   when   linking  the   observed
properties  of galaxies  to those  of  their dark  matter halos.   Our
results imply that great care is  to be taken in linking an observable
velocity  such as $V_{\rm  max}$ to  the circular  velocity of  a dark
matter halo.  Based  on our results, we suggest  that the construction
of a halo {\it mass}  function using $M_{\rm vir} \propto R_d V^2_{\rm
max}$ may proof  more reliable. More details of  the results presented
here can be found in van den Bosch (2002b)

%


\begin{thebibliography}{8.}
\addcontentsline{toc}{section}{References}

\bibitem{BdJ01} E. F. Bell, R. S. de Jong: ApJ, \textbf{550}, 212 (2001)

\bibitem{Bul01} J. S. Bullock et al.: ApJ, \textbf{555}, 240 (2001)

\bibitem{FA00} C. Firmani, V. Avila-Reese: MNRAS, \textbf{315}, 457 (2000)

\bibitem{Gon00} A. H. Gonzales et al.: ApJ, \textbf{528}, 145 (2000)

\bibitem{Koc01} C. S. Kochanek: preprint, astro-ph/0108160 (2001)

\bibitem{MMW98} H. J. Mo, S. Mao, S. D. M. White: MNRAS, \textbf{295},
319 (1998)

\bibitem{NFW97} J. F. Navarro, C. S. Frenk, S. D. W. White: ApJ, 
\textbf{490}, 493 (1997)

\bibitem{New00} J. A. Newman, M. Davis: ApJ, \textbf{534}, L11 (2000)

\bibitem{vdB98} F. C. van den Bosch: ApJ, \textbf{507}, 601 (1998)

\bibitem{vdB00} F. C. van den Bosch: ApJ, \textbf{530}, 177 (2000)

\bibitem{vdB01} F. C. van den Bosch: MNRAS, \textbf{327}, 1334 (2001)

\bibitem{vdB02a}   F.    C.   van   den   Bosch:   MNRAS,  in   press;
astro-ph/0105158 (2002a)

\bibitem{vdB02b} F.  C.  van den Bosch: MNRAS, submitted;
astro-ph/0112566 (2002b)

\end{thebibliography}
\end{document}